\documentclass[aps,twocolumn, prl]{revtex4}
\usepackage{graphicx}
\usepackage[scanall]{psfrag}
\usepackage{amssymb}

\begin{document}

%
%
%
%
\def\oti{{\otimes}}
\def\lb{ \left[ }
\def\rb{ \right]  }
\def\tilde{\widetilde}
\def\bar{\overline}
\def\hat{\widehat}
\def\*{\star}
\def\[{\left[}
\def\]{\right]}
\def\({\left(}		\def\BL{\Bigr(}
\def\){\right)}		\def\BR{\Bigr)}
	\def\BBL{\lb}
	\def\BBR{\rb}
%
%
\def\zb{{\bar{z} }}
\def\zbar{{\bar{z} }}
\def\frac#1#2{{#1 \over #2}}
\def\inv#1{{1 \over #1}}
\def\half{{1 \over 2}}
\def\d{\partial}
\def\der#1{{\partial \over \partial #1}}
\def\dd#1#2{{\partial #1 \over \partial #2}}
\def\vev#1{\langle #1 \rangle}
\def\ket#1{ | #1 \rangle}
\def\rvac{\hbox{$\vert 0\rangle$}}
\def\lvac{\hbox{$\langle 0 \vert $}}
\def\2pi{\hbox{$2\pi i$}}
\def\e#1{{\rm e}^{^{\textstyle #1}}}
\def\grad#1{\,\nabla\!_{{#1}}\,}
\def\dsl{\raise.15ex\hbox{/}\kern-.57em\partial}
\def\Dsl{\,\raise.15ex\hbox{/}\mkern-.13.5mu D}
%
%
\def\ga{\gamma}		\def\Ga{\Gamma}
\def\be{\beta}
\def\al{\alpha}
\def\ep{\epsilon}
\def\vep{\varepsilon}
\def\la{\lambda}	\def\La{\Lambda}
\def\de{\delta}		\def\De{\Delta}
\def\om{\omega}		\def\Om{\Omega}
\def\sig{\sigma}	\def\Sig{\Sigma}
\def\vphi{\varphi}

%
%
\def\CA{{\cal A}}	\def\CB{{\cal B}}	\def\CC{{\cal C}}
\def\CD{{\cal D}}	\def\CE{{\cal E}}	\def\CF{{\cal F}}
\def\CG{{\cal G}}	\def\CH{{\cal H}}	\def\CI{{\cal J}}
\def\CJ{{\cal J}}	\def\CK{{\cal K}}	\def\CL{{\cal L}}
\def\CM{{\cal M}}	\def\CN{{\cal N}}	\def\CO{{\cal O}}
\def\CP{{\cal P}}	\def\CQ{{\cal Q}}	\def\CR{{\cal R}}
\def\CS{{\cal S}}	\def\CT{{\cal T}}	\def\CU{{\cal U}}
\def\CV{{\cal V}}	\def\CW{{\cal W}}	\def\CX{{\cal X}}
\def\CY{{\cal Y}}	\def\CZ{{\cal Z}}

\def\rvac{\hbox{$\vert 0\rangle$}}
\def\lvac{\hbox{$\langle 0 \vert $}}
\def\comm#1#2{ \BBL\ #1\ ,\ #2 \BBR }
\def\2pi{\hbox{$2\pi i$}}
\def\e#1{{\rm e}^{^{\textstyle #1}}}
\def\grad#1{\,\nabla\!_{{#1}}\,}
\def\dsl{\raise.15ex\hbox{/}\kern-.57em\partial}
\def\Dsl{\,\raise.15ex\hbox{/}\mkern-.13.5mu D}
%
%
%
\font\numbers=cmss12
\font\upright=cmu10 scaled\magstep1
\def\stroke{\vrule height8pt width0.4pt depth-0.1pt}
\def\topfleck{\vrule height8pt width0.5pt depth-5.9pt}
\def\botfleck{\vrule height2pt width0.5pt depth0.1pt}
\def\Zmath{\vcenter{\hbox{\numbers\rlap{\rlap{Z}\kern
0.8pt\topfleck}\kern 2.2pt
                   \rlap Z\kern 6pt\botfleck\kern 1pt}}}
\def\Qmath{\vcenter{\hbox{\upright\rlap{\rlap{Q}\kern
                   3.8pt\stroke}\phantom{Q}}}}
\def\Nmath{\vcenter{\hbox{\upright\rlap{I}\kern 1.7pt N}}}
\def\Cmath{\vcenter{\hbox{\upright\rlap{\rlap{C}\kern
                   3.8pt\stroke}\phantom{C}}}}
\def\Rmath{\vcenter{\hbox{\upright\rlap{I}\kern 1.7pt R}}}
\def\Z{\ifmmode\Zmath\else$\Zmath$\fi}
\def\Q{\ifmmode\Qmath\else$\Qmath$\fi}
\def\N{\ifmmode\Nmath\else$\Nmath$\fi}
\def\C{\ifmmode\Cmath\else$\Cmath$\fi}
\def\R{\ifmmode\Rmath\else$\Rmath$\fi}

\def\barray{\begin{eqnarray}}
\def\earray{\end{eqnarray}}
\def\beq{\begin{equation}}
\def\eeq{\end{equation}}

\def\no{\noindent}
\def\chidag{\chi^\dagger}
\def\Det{{\rm Det}}
\def\Tr{{\rm Tr}}
\def\det{{\rm det}}
\def\tr{{\rm tr}}
\def\Upp{ U''}
\def\Svec{\vec{S}}
\def\mvec{\vec{m}}
\def\zdag{z^\dagger}
\def\sigmavec{\vec{\sigma}}
\def\sigvec{\sigmavec}
\def\svec{{\vec{n}\,}}
\def\mvec{\svec}
\def\lambdastar{\lambda_*}
\def\dim#1{\lbrack\!\lbrack #1 \rbrack\!\rbrack }
\def\chidagger{\chi^\dagger}
\def\nvec{\vec{n}}
\def\OM#1#2{O_{#1}^{(#2)}}
\def\Sp#1#2{Sp_{#1}^{(#2)}}

\def\xvec{\vec{x}}

\title{Quantum critical  spin liquids 
and superconductivity in the cuprates}
\author{Andr\'e  LeClair}
\affiliation{Newman Laboratory, Cornell University, Ithaca, NY} 
\date{October 2006}

\bigskip\bigskip\bigskip\bigskip

\begin{abstract}

We describe a new kind of quantum critical point in 
the context of quantum anti-ferromagnetism  in $2d$ 
that can be understood as  a quantum critical spin liquid.    
Based on the comparison of exponents with previous numerical work,  
we argue it
describes a  transition from an anti-ferromagnetic
N\'eel ordered state to a VBS-like state.   We
argue further that the symplectic fermions capture
the proper degrees of freedom in the zero temperature
phase that is the parent to the superconducting phase in
the cuprates.   
We then show that our model reproduces some
features found recently in experiments  and also
in the Hubbard model.

\end{abstract}

\maketitle

\section{Introduction}

Over the last few years there has been much interest
in finding new quantum critical points in the context
of quantum anti-ferromagnetism in $2d$.   One motivation
are their possible applications to the anti-ferromagnetic
phase of the Hubbard model and to superconductivity in
the cuprates.    Strong arguments were given by
Senthil et. al. that there should exist a quantum
critical point that separates a N\'eel order 
phase from a valence-bond solid like phase\cite{Senthil}.  
In the model these authors considered,  it appears  difficult
to exhibit  this critical point perturbatively.   In this
paper we consider a different, simpler model   which contains a 
critical point that can easily be studied with a perturbative
renormalization group analysis.   Our  critical
exponents agree very favorably with the numerical
simulation of the model in \cite{Senthil} carried
out by Motrunich and Vishwanath\cite{Motrunich},
and we interpret this as strong evidence that our model
is in the same universality class.    In the second part of
the paper we take some initial steps toward applying this
theory to superconductivity.

For the remainder of this Introduction, 
we summarize the motivations given in \cite{Andre1}
for our model.   It is well-known that the continuum limit of the
Heisenberg anti-ferromagnet constructed over 
a N\'eel ordered state  leads to a non-linear sigma model
for a 3-component field $\nvec (\xvec)$ satisfying
$\nvec^2 = 1$.  (For a detailed account of this in
$1d$ and $2d$, with additional references to the original
works see\cite{Fradkin}.)   In $1d$ a topological term $S_\theta$
arises directly in the map to the continuum and affects
the low-energy (infra-red (IR)) limit: half-integer
spin chains are gapless whereas the integer ones are gapped. 
This is the well-known  Haldane conjecture\cite{Haldane}.  
It is known from the exact Bethe-ansatz solution of 
the spin $1/2$ chain\cite{Bethe}  that
the low-lying excitations are spin $1/2$ particles 
referred to a spinons\cite{Faddeev}.

In $2d$ one still obtains a non-linear sigma model.
The topological term  $S_\theta$ also arises but, unlike
in $1d$,  it is a  renormalization group (RG) irrelevant operator
so we discard it.   The non-linear constraint $\nvec^2 =1$ 
renders the model non-renormalizable in $2d$.  However 
there is a quantum critical point in the spin system\cite{Halperin,Ye}
that is captured by the following euclidean space action: 
\beq
\label{I.4}
S_{WF}  =  \int d^3 x \( \inv{2} \d_\mu \mvec \cdot \d_\mu \mvec  +~~
\tilde{\lambda} \,  (\mvec\cdot \mvec )^2 \)
\eeq
$\d_\mu^2=  \sum_{\mu=1}^3  \d_{x_\mu}^2$.  
The fixed point is in the Wilson-Fisher univerality class\cite{Wilson}.
The fixed point generalizes to 
an $M$ component vector $\nvec$ and 
we  will refer to the fixed point theory as $\OM{M}{D}$ 
where $D=d+1$.

Senthil et. al.\cite{Senthil} have given numerous arguments
suggesting that anti-ferromagnets can have more exotic
quantum critical points that are not in the Wilson-Fisher
universality class.   They are expected to describe 
for instance transitions between a N\'eel ordered state
and a valence-bond-solid (VBS)-like phase, and some evidence
for such a transition was found by Motrunich and Vishwanath\cite{Motrunich}. 
See also \cite{Sandvik}. 
A large part of the literature devoted to the ``deconfined'' quantum
critical points represent the $\nvec$ field as 
\beq
\label{I.5}
\mvec =  \chidag  \vec{\sigma} \chi 
\eeq
where $\sigmavec$ are the Pauli matrices and $\chi = (\chi_1 , \chi_2)=
\{ \chi_i  \} $ is a two component complex {\it bosonic}  spinor.
The constraint $\mvec^2 =1$ then follows from   the constraint
$\chidag \chi  = 1$.   Coupling $\chi $ to a $U(1)$ gauge field $A_\mu$ 
with the covariant derivative $D_\mu = \d_\mu - i A_\mu$, then by 
eliminating the non-dynamical gauge field using it's equations of motion,
one can show that  the following actions are
equivalent:
\beq
\label{I.6}
{\textstyle \inv{2}}
\int d^{D} x ~  
 \d_\mu \mvec \cdot \d_\mu \mvec = \int d^{D} x 
~| D_\mu \chi  |^2  
\eeq
Senthil et. al. considered a model where $\chi$ was a boson,
and added an $F_{\mu \nu}^2$ term which makes the gauge field
dynamical.

The central idea of this paper is that the spinon $\chi$ 
is a {\it fermion}.    Numerous arguments were given in\cite{Andre1}.
First of all, the equivalence  (\ref{I.6}) is valid whether
$\chi$ is a boson or fermion.  
Secondly, suppose the theory is asymptotically free in the ultra-violet, 
which it is.  
Then in this conformally invariant limit, one would hope
that the description in terms of $\nvec$ or $\chi$ 
have the same numbers of degrees of freedom.   One way to
count these degrees of freedom is to compute the free energy
density at finite temperature.  For a single species of massless
particle, the free energy density in $2d$ is 
\beq
\label{free.2}
\CF = - c_3 \frac{\textstyle \zeta(3)}{2\pi} T^3
\eeq
where $c_3=1$ for a boson and $3/4$ for a fermion.   
($\zeta$ is Riemann's zeta function.)  
(In $1d$ the analog of the above is $\CF = - c \pi T^2/6$, where 
$c$ is the Virasoro central charge\cite{Cardy,Affleck}.)  
Therefore one sees that the 3 bosonic degrees of freedom 
of an $\svec$ field  has the same $c_3$ as an $N=2$  component
$\chi$ field.     One way to possibly  understand the
change of statistics from bosonic to fermionic is
by simply adding a Chern-Simons term to (\ref{I.6})\cite{Wilczek}.  

 In analogy with the bosonic non-linear sigma
model,  since the constraint $\chidag \chi =1$ again renders
the model non-renormalizable,  we relax this constraint and 
consider the action:
\beq
\label{central.1}
S_\chi =  \int d^D x ~  \( 2 \d_\mu \chi^\dagger \d_\mu \chi 
 + 16 \pi^2 \lambda 
 ~ |\chi^\dagger \chi|^2 \)
\eeq
where now $\chi$ is an $N$-component complex field, sometimes
referred to as a symplectic fermion, $\chidag\chi = \sum_{i=1}^N
\chidag_i \chi_i $.   
This model may at first appear peculiar, in that 
the field $\chi$ has a Klein-Gordon action but is quantized
as a fermion.  However, we remind the reader that there is
no spin-statistics theorem in $2d$.  
Note there is no gauge field (``emergent photon'') in the model.

As shown in \cite{Andre1},  the $(\chidag\chi)^2$ interactions 
drive the theory to a new  infrared stable fixed point, we
refer to as $\Sp{N}{D}$.  For $N=2$ in $3D$  
the exponents were computed to be\cite{Andre1} 
\beq
\label{exps}
\eta = 3/4, ~~~\nu = 4/5, ~~~\beta = 7/10,~~~\delta = 17/7  
~~~~(\Sp{2}{3})
\eeq
(The definition of these exponents is given in the next section.)
These  agree very favorably with the critical exponents found
in \cite{Motrunich}: 
$~~\nu = .8\pm 0.1, ~ \beta / \nu  = .85 \pm 0.05$, certainly
within error bars.     The shift down to $3/4$ 
from the classical value $\eta = 1$ is entirely due to the
fermionic nature of the $\chi$ fields. 
We thus conjecture that the $\Sp{2}{3}$ model describes
a deconfined quantum critical spin liquid.

In the next section we summararize the results of the critical
theory found in \cite{Andre1}.  In section III 
we apply this model to high $T_c$ superconductors and
describe agreement  with some recent experimental results\cite{Seamus}.

\section{The critical theory}

The 1-loop  beta function for the $N$ component model in $D$ space-time
dimensions is 
\beq
\label{fp.3}
\frac{d\lambda}{d\ell} = (4-D) \lambda + (N-4) \lambda^2
\eeq
where increasing the length 
$e^\ell$ corresponds to the flow toward low energies. 
The above beta function has a zero at 
\beq
\label{fp.4}
\lambdastar = \frac{4-D}{4-N}
\eeq
Note that $\lambda_*$ changes sign at $N=4$.  
It is not necessarily a problem to have a fixed point at
negative $\lambda$ since the particles are fermionic:
the energy is not unbounded from below because of the
Fermi sea.   Near $\lambda_*$ one has that
$d \lambda /d\ell \sim (D-4)(\lambda - \lambdastar)$
which implies the fixed point is IR stable regardless of  
the sign of $\lambda_*$,  so long as $D<4$.
Arguments were given in \cite{Andre1} that at $N=4$, 
two of the $\chi$ components reconfine into a spin field $\nvec$,
though we will not need this here.

\def\xvec{{\bf x}}
\def\yvec{{\bf y}}
\def\zvec{{ \bf z}}
\def\gammachi{{\gamma_\chi}}
\def\gammam{{\gamma_m}}
\def\absx{|\xvec|}

\subsection{Definition of the exponents for the $\nvec$ field }

Though the spinons $\chi$ are deconfined, it
is still physically meaningful to define exponents
in terms of the original order parameter $\nvec$,
which is represented by eq. (\ref{I.5}).
These exponents are especially useful if
one approaches the fixed point from within 
an anti-ferromagnetic phase.  
We then define the exponent $\eta$ as the one
characterizing the spin-spin correlation function:
\beq
\label{scale.1}
\langle \svec (\xvec ) \cdot \svec (0) \rangle \sim 
\inv{|\xvec|^{D-2 + \eta}}
\eeq
For the other exponents we need a measure of the
departure from the critical point;  these are the
parameters that are tuned to the critical point in
simulations and experiments:
\beq
\label{scale.2}
S_\chi \to S_\chi + \int d^D x ~(m^2 \, \chidag \chi  + 
\vec{B}\cdot \nvec  )
\eeq
Above, $m$ is a mass and $\vec{B}$ the magnetic field.
The correlation length exponent $\nu$,   and magnetization 
exponents $\beta, \delta$ 
are then defined by
\beq
\label{scale.3}
\xi \sim m^{-\nu}, ~~~~~\langle \svec \rangle \sim m^\beta \sim 
B^{1/\delta} 
\eeq
Above $\langle \nvec \rangle$ is the one-point function of
the field $\nvec ({\bf x})$ and is independent of ${\bf x}$ 
by the assumed translation invariance.

The above exponents are related to the anomalous dimension of
the field $\chi$ and the operator $\chidag \chi$.   This leads
to the following relations amoung the exponents:
\beq
\label{betaeta}
\beta = \nu(D-2 + \eta)/2, ~~~~ 
\delta = \frac{D+(2-\eta)}{D -(2- \eta )}
\eeq
The lowest order contributions to the anomalous dimensions of
the operator $\chidag \chi$ arise at 1-loop, and for $\chi$
at two loops. 
The calculation in \cite{Andre1} gives in $3D$: 
\beq
\label{statmech.1}
\nu = \frac{2 (4-N)}{7-N} , ~~~~~~
\beta = \frac{2N^2 - 17N + 33}{N^2 - 11N + 28}
\eeq
For $N=2$ one obtains the results quoted in the introduction.  

It was conjectured in\cite{Andre1}  that the
$\Sp{-N}3$ model has the same fixed point as the $\OM{N}3$ model,
so that $\Sp{-1}3$ is the $3D$ Ising model.  The exponents
are in very good agreement with known Ising exponents.  
A shorter version of these results was  described in\cite{Andre2}.

\section{Superconductivity based on symplectic fermions}

In this section we explain  how our quantum critical spin
liquid could be relevant to  the 
understanding of superconductivity in the cuprates, 
which is believed to be a $2+1$ dimensional problem\cite{Anderson2}.  
To do this, one must turn  to the language of the Hubbard
model.   In the anti-ferromagnetic phase of the Hubbard model,
the spin field $\nvec = c^\dagger \sigmavec c $,  where
$c$ are the physical electrons.   Therefore in applying our
model to the Hubbard model,  the symplectic fermion $\chi$ 
is a descendant of  the electron, so  it can  carry electric charge.  
Consider the zero temperature phase diagram of the cuprates as a function of
the density of holes.   At low density there is an anti-ferromagnetic
phase.  Suppose that  the first quantum critical point
is a transition from a N\'eel ordered 
to a VBS-like  phase and is well described by our
symplectic fermion model at $N=2$.  Compelling
evidence for a VBS like phase has recently been
seen by Davis' group\cite{Seamus};  and it in fact resembles
more a ``VBS spin glass''.   The superconducting
phase actually originates from this VBS-like phase.    
It is then possible that the 2-component 
$\chi$ fields capture
the correct degrees of freedom for the description of this 
VBS-like phase.   These fermionic spin $1/2$ spinon quasi-particles
acquire a gap away from the critical point,
which is described by the mass term in
eq. (\ref{scale.2}).  Note that away from the quantum
critical point, the particles already have a gap $m$
because of the relativistic nature of the symplectic
fermion\footnote{Here the ``speed of light'' is some
material dependent quantity such as a Fermi velocity. 
We set it to 1.}.

\def\adag{a^\dagger}
\def\bdag{b^\dagger} 
\def\kvec{{\bf k}}

Superconductivity based on the symplectic fermion has some
very interesting features.      
In the VBS-like  phase the $\chi$-particles are charged
fermions and it's possible that additional phonon interactions,
or even the $\chi^4$ interactions that led to the critical theory,
could
lead to a pairing interaction that causes them to condense
into Cooper pairs just as in the usual BCS theory.   
Recent numerical work on the Hubbard model suggests that
the Hubbard interactions themselves can provide
a pairing mechanism\cite{Scalapino}.  

 Passing
to Minkowski space,  the hamiltonian of the symplectic
fermion is
\barray
\nonumber
H &=& \int d^2 \xvec  \Bigl(  2 \, \d_t \chi^\dagger \d_t \chi + 
2 \vec{\nabla} \chi^\dagger \cdot \vec{\nabla} \chi
\\ \label{highT.1} 
&~& ~~~~~~~~~~~~~~+ m^2 \chidag \chi  +  \tilde{\lambda} 
~ (\chidag \chi)^2 \Bigr)
\earray
Expand the field in terms of creation/annihilation operators
as follows
\barray
\label{highT.2}
\chi (\xvec ) &=& \int \frac{ d^2 \kvec }{4\pi \sqrt{\omega_{\kvec}}} 
\( a_\kvec \, e^{-i\kvec \cdot \xvec} + b_\kvec \, e^{i \kvec\cdot \xvec} \)
\\ \nonumber
\chidag  (\xvec ) &=& \int \frac{ d^2 \kvec }{4\pi \sqrt{\omega_{\kvec}}} 
\( \adag_\kvec \, 
e^{i\kvec \cdot \xvec} + \bdag_\kvec \, e^{-i \kvec\cdot \xvec} \)
\earray
where $\omega_\kvec = \sqrt{\kvec^2 + m^2 }$. 
Canonical quantization of the $\chi$-fields, 
$\{ \chidag (\xvec) , \d_t \chi (\xvec' )\} = i \delta(\xvec - \xvec' )/2$,
leads to the anti-commutation relations 
\beq
\label{hight.3}
\{ \bdag_\kvec , b_{\kvec'} \} = - \{ \adag_\kvec , a_{\kvec'} \} = 
\delta_{\kvec , \kvec'} 
\eeq
The free hamiltonian is then
\beq
\label{highT.4}
H_0 = \int d^2 \kvec ~  \omega_\kvec \,  
\( \adag_\kvec a_\kvec + \bdag_\kvec b_\kvec \) 
\eeq

The minus sign in the anti-commutator of the $a$'s means there
are negative norm states in the free Hilbert space\footnote{I thank
S. Sachdev for pointing out this potential difficulty.}.      However
a simple projection onto even numbers of $a$-particles gives
a unitary Hilbert space.  In a potential physical realization,
since the anti-ferromagnetic spin field $\nvec$ is deconfined,
it is clear that the particles come in pairs.

The minus sign in eq. (\ref{hight.3}) actually leads to  a two-band
theory.  
This has been seen experimentally\cite{Seamus} and
also in the Hubbard model\cite{Scalapino}.
There are two kinds of spin 1/2 particles created by $a$ or $b$:
$\adag_\kvec |0\rangle = |\kvec\rangle_a , ~~
\bdag_\kvec | 0 \rangle = | \kvec \rangle_b$,
with energies $\vep_{a,b}$: 
\barray
\label{highT.5}
H_0 \, | \kvec \rangle_{a,b} &=&
 \vep_{a,b} (\kvec ) \, | \kvec  \rangle_{a,b}  
\\ \nonumber
\vep_b (\kvec ) = \omega_\kvec , ~&~&~
\vep_a (\kvec) = - \omega_\kvec   
\earray

\def\omegak{\omega_\kvec}

Note that $\vep_a \leq -m$ and $\vep_b \geq m$ 
so there is a gap $2m$.  The density of states can easily be
computed in the free theory.  
The density of states per volume is defined 
so that  
\beq
\label{density.0}
n = \int \frac{d^2 \kvec}{(2\pi)^2} ~ \rho (\kvec ) =  
\int d\vep ~ \rho (\vep) 
\eeq
where $n$ is the particle number density. 
Using    
$\inv{(2\pi)^2} \int d^2 \kvec  = \int d\vep \, \vep / 2\pi$, 
one
finds
\barray
\label{density.2}
\rho (\vep ) &=& 
\frac{\vep}{2\pi} \, 
f_b (\vep) ~~~~~ {\rm for} ~ \vep \geq m 
\\ \nonumber
&=& 0 ~~~~~~~~~~~~~~~{\rm for~}    -m < \vep < m 
\\ \nonumber 
 &=& 
\frac{\vep}{2\pi} \, 
 f_a (\vep)  ~~~~~ {\rm for~} \vep \leq - m
\earray
where $f_{a,b}$ are temperature dependent Fermi-Dirac
filling fractions.     Interactions will tend to 
fill the gap.

The last ingredient one needs is a  pairing phase transition,
so let us turn to the interactions.  
The $(\chidag \chi)^2$ interaction
is very short ranged since it corresponds to a $\delta$-function 
potential in position space.  Because of the relativistic nature
of the fields, the interaction gives rise to a variety of 
pairing interactions.    There are actually pairing
interactions between the two 
bands.  However let us focus on the pairing interactions
within each band that resemble BCS pairing.
If  all momenta have roughly the same magnitude
$|\kvec|$, then the interaction gives the terms  (up to factors of $\pi$):
\beq
\label{highT.6}
H_{\rm int} =   - \tilde{\lambda} \sum_{\kvec, ~i,j = \uparrow, \downarrow}  
\( \adag_{\kvec , i} \adag_{-\kvec , j} a_{-\kvec , i} a_{\kvec , j} 
 + (a\to b) \) + ....
\eeq
The overall minus sign of the interaction is due to a fermionic
exhange statistics.  
Because of the overall minus sign this is an attractive 
pairing interaction as in BCS.    One difference is that
in addition to the opposite spin pairing
interactions with $i\neq j$,  there are also equal-spin
pairings.   The quantum ground state can be further studied
by reasonably straightforward application of the mean-field BCS
construction\cite{Schrieffer}.

\section{Conclusions}

We have shown that 
the 2-component relativistic symplectic fermion appears to have
some of the  right ingredients
to  explain the zero temperature phase diagram of 
the high $T_c$ cuprates.  It has a quantum critical point 
that we have interpreted as a transition between 
an anti-ferromagnetic phase and VBS-like phase.   
Away from the critical point the quantum spin liquid has a 
2-band structure  as  in the VBS spin-glass
phase\cite{Seamus}.   It also naturally has BCS-like pairing
interactions.    A real test  of our model would
be a measurement of the critical properties of the 
anti-ferromagnetic to VBS spin-glass phase.  The magnetic
exponent $\delta$  is probably the easiest to measure and 
our theory predicts $\delta = 17/7$.

\section{Acknowledgments}

I would especially like to thank Seamus Davis for explaining
his most recent results before publication, 
and C. Henley, E. Mueller, S. Sachdev, T. Senthil, and
Jim Sethna for discussions.

\vspace{.4cm}
\end{document}